\documentclass[11pt]{article}
\pdfoutput=1
\usepackage{epsfig}\parskip 5pt plus 1pt
\usepackage{bbm}
\usepackage{amsmath}
\usepackage{amssymb}
\usepackage{amsfonts}
\usepackage{mathrsfs}
\usepackage{graphicx}
\usepackage{color}
\newcommand{\be}{\begin{equation}}
\newcommand{\ee}{\end{equation}}
\newcommand{\bea}{\begin{eqnarray}}
\newcommand{\eea}{\end{eqnarray}}

\begin{document}
{\vspace*{-2cm}
\flushright{ULB-TH/08-35}
\vspace{-4mm}
\vskip 1.5cm}
\begin{center}
{\Large \bf Hidden vector dark matter}
\end{center}
\vskip 0.5cm
\begin{center}

{\large Thomas Hambye}\footnote{thambye@ulb.ac.be}
\\
\vskip .5cm
Service de Physique Th\'eorique,\\
\vspace{0.7mm}
Universit\'e Libre de Bruxelles, 1050 Brussels, Belgium\\

\end{center}
\vskip 0.5cm

\begin{abstract}

We show that dark matter could be made of massive gauge bosons whose stability doesn't require to impose by hand any discrete or global symmetry. Stability of gauge bosons can be guaranteed by the custodial symmetry associated to the gauge symmetry and particle content of the model. The particle content we consider to this end is based on a hidden sector made of a vector multiplet associated to a non-abelian gauge group and of a scalar multiplet charged under this gauge group. The hidden sector interacts with the Standard Model particles through the Higgs portal quartic scalar interaction in such a way that the gauge bosons behave as thermal WIMPS. This can lead easily to the observed dark matter relic density in agreement with the other various  constraints, and can be tested experimentally in a large fraction of the parameter space. In this model the dark matter direct detection rate and the annihilation cross section can decouple if the Higgs portal interaction is weak.
\end{abstract}
\setcounter{footnote}{0}
\vskip2truecm

\section{Introduction}
In the Standard Model (SM) the best known sector is the gauge sector. Besides Lorentz symmetry and CPT symmetry, all fundamental symmetries of the SM are gauge symmetries. There are also accidental symmetries, in particular $B-L$ number conservation (or $B$ and $L$ separately neglecting highly suppressed instanton effects), but also this one is closely related to the gauge symmetries. It results from the particle content of the SM and the gauge symmetry charges assigned to it. There are no ad-hoc discrete or global symmetries. This leads to a certain number of stable particles, the photon (stable due a conserved gauge symmetry which makes it massless), the electron (stable because it is the lightest particle charged under this conserved gauge symmetry), the lightest neutrino (due to Lorentz invariance and the fact that it is the lightest fermion) and the proton (from the accidental baryon number conservation). Any new physics model must not spoil the stability of these particles with huge accuracy.

In the following we assume that this gauge symmetry reason for having stable particles also holds for the dark matter (DM) particle. Of course there is no mandatory reason why this must be the case for DM too, but the fact that this holds for the SM is puzzling enough to investigate this possibility.
Along this line we consider a particularly simple model where DM is made of a multiplet of vector particles communicating with the SM through the Higgs portal, and show it is a perfectly viable DM candidate.\footnote{Our perspective is purely phenomenological, and therefore different from the top-down one of particular grand-unified
supersymmetric models which can achieve the same goal, i.e.~obtaining R-parity conservation as a remnant symmetry of a spontaneously broken gauge group, e.g.~gauged $U(1)_{B-L}$  \cite{Martin:1992mq}.}
In the process we present a new model with spin-1 DM. Most DM models involve particles
of spin 0 (scalar singlet \cite{singletDM,Barger:2007im,Andreas:2008xy}, axion \cite{axion}, inert 
doublet \cite{inertDM,Andreas:2008xy}, high $SU(2)_L$ scalar multiplets \cite{Cirelli:2005uq},...), spin $1/2$ (neutralino \cite{neutralino}, axino \cite{axino}, high $SU(2)_L$ fermion multiplets \cite{Cirelli:2005uq}, singlet fermion \cite{Pospelov,Belanger},....) or spin $3/2$ (gravitino \cite{gravitino}) particles. To our knowledge models of spin-1 DM which have been proposed involve either more than 4 dimensions \cite{Servant:2002aq} or an explicit discrete symmetry (little Higgs DM with T-parity~\cite{littleHiggs}).


\section{Model}

If there were no fermions
in the SM, and no mixing between $SU(2)_L$ and $U(1)_Y$, the $W$ and $Z$ bosons would be degenerate in mass and stable due to the custodial symmetry of the scalar and $SU(2)_L$ gauge SM sectors. As, fortunately,
there are fermions (and $SU(2)_L\times U(1)$ mixing), 
the $W$ and $Z$ are not stable, but this possibility to have stable gauge bosons could hold in a hidden sector coupling to the SM through the Brout-Englert-Higgs (Higgs for short) portal quartic scalar interaction.

The scenario is based on two simple assumptions. First we assume the existence of a $A'^\mu$ gauge multiplet associated to a non-abelian gauge symmetry, $G'$. Under this gauge group all SM particles are singlets. Second this non-abelian field couples to the SM only through a complex scalar Higgs portal field, $\phi$, which is singlet of the SM but charged under $G'$. Mixing of $A'_\mu$ with the SM gauge bosons (through $F_{\mu \nu} F^{\mu \nu}_Y$ kinetic mixing with the hypercharge gauge field) is forbidden by the non-abelian character of the extra gauge symmetry.
For instance we consider the simple case $G'=SU(2)$, which we denote $SU(2)_{HS}$, and we take $\phi$ to be a complex doublet of this gauge symmetry. No discrete symmetry are assumed at any level. 
The most general Lagrangian one can write with these simple assumptions is
\be
{\cal L}= {\cal L}^{SM} -\frac{1}{4} F'^{\mu\nu} \cdot F'_{\mu \nu}
+(D_\mu \phi)^\dagger (D^\mu \phi) -\lambda_m \phi^\dagger \phi H^\dagger H-\mu^2_\phi \phi^\dagger \phi -\lambda_\phi (\phi^\dagger \phi)^2 \,,
\label{inputlagr}
\ee
with $D^\mu \phi=\partial^\mu \phi - i\frac{g_\phi}{2} \tau \cdot A'^\mu$.  In the SM lagrangian we define the Higgs potential notations as: ${\cal L}^{SM} \owns -\mu^2 H^\dagger H -\lambda (H^\dagger H)^2$ with $H=(H^+,H^0)$.

If $SU(2)_{HS}$ is spontaneously broken (i.e.~$\mu^2_\phi<0$), writing as usual, 
$\phi=exp(i \tau \cdot\xi/v_\phi) \cdot (0, \,\frac{1}{\sqrt{2}} [v_\phi+\eta'] )^T$,
and gauge rotating away the $\xi$ part to absorb it in $A_{\mu}=U A'_\mu U^{-1}-\frac{i}{g} [\partial_\mu U] U^{-1}$ with $U=exp(-i\tau \cdot \xi/v_\phi)$
we get
\begin{eqnarray}
{\cal L}&=&{\cal L}_{SM}-\frac{1}{4} F_{\mu \nu} \cdot F^{\mu \nu} +\frac{1}{8} (g_\phi v_\phi)^2 A_\mu \cdot A^\mu+\frac{1}{8} g_\phi^2 A_\mu \cdot A^\mu \eta'^2 +\frac{1}{4} g_\phi^2 v_\phi A_\mu \cdot A^\mu \eta' \nonumber\\
&&+\frac{1}{2}(\partial_\mu \eta')^2 -\frac{\lambda_m}{2}(\eta'+v_\phi)^2 H^\dagger H- \frac{\mu_\phi^2}{2} (\eta'+v_\phi)^2 - \frac{\lambda_\phi}{4} (\eta'+v_\phi)^4 \,,
\label{lagrzerov}
\end{eqnarray}
which gives $m_{A}=g_\phi v_\phi/2$ and $m^2_{\eta'}=-2 \mu_\phi^2$. 

The Lagrangian above has an important property: since the scalar field is in the fundamental representation of the gauge group it displays a custodial symmetry, $SO(3)$ in the $A^\mu_{1,2,3}$ component space. As a result the 3 $A^\mu_i$ components are degenerate in mass and are stable. Since all other particles (as $\eta'$) are $SO(3)$ singlets, any decay of the gauge bosons is forbidden by the custodial symmetry. In practice, this stability can be also seen from the facts that:
\begin{itemize}
\item All interactions from the scalar kinetic term involve the $A^\mu$ field in pairs. The
dangerous $d^\mu \phi A^\mu_i$ terms which would make the gauge boson instable, disappear from the absorption of the Goldstone bosons by the gauge bosons, as in the SM. These scalar kinetic term interactions do not mix the various $A^\mu_i$ $SU(2)_L$ components. 
\item In the pure gauge sector there are quadrilinear and trilinear interactions but they do not cause any decay of the gauge boson fields. The quadrilinear terms because they involve $A^\mu_i$ components in pairs. The trilinear terms because they involve three different $A^\mu_i$ components - they are of the form $\varepsilon_{ijk} \partial_\mu A_{i \nu} (A^\mu_j A^\nu_k-A^\nu_j A^\mu_k)$ - while all other interactions, which would allow the vector bosons to decay (to scalars), involve two same components.

\end{itemize}

This custodial symmetry structure is closely associated to the particle content of the model. 
For instance, light fermions charged under $G'$ or higher scalar multiplets of $G'$ would make the gauge bosons unstable, see section 8. But in a similar way could we think about new particles destabilizing the proton which
nevertheless turns out to be stable with huge accuracy.

In order to get the mass spectrum of this model it is necessary to minimize the potential not only along the $\phi$ direction as above but also along the $H$ direction. Writing $H=exp(i \tau \cdot \zeta/v)(0, \,\frac{1}{\sqrt{2}} [v+h'] )^T$, with $v=246$~GeV, going to the unitary gauge and imposing $d V/dh=dV/d\eta=0$ we get:
\begin{eqnarray}
v^2&=&\frac{-\mu^2 \lambda_\phi +\frac{1}{2}\lambda_m \mu^2_\phi}{\lambda \lambda_\phi-\frac{1}{4} \lambda_m^2} \,,\\
v_\phi^2&=&\frac{-\mu_\phi^2 \lambda +\frac{1}{2}\lambda_m \mu^2}{\lambda \lambda_\phi-\frac{1}{4} \lambda_m^2} \,.
\end{eqnarray}
This leads to a non-diagonal mass matrix in the $(h', \,\eta')$ basis
\begin{equation}
M^2=\left(
\begin{array}{ cc}
   m^2_{h'}&  m^2_{h' \eta'}\\
    m^2_{h' \eta'} & m^2_{\eta'}
\end{array}
\right) \,,\,
\label{massmatrix}
\end{equation} 
with
\begin{eqnarray}
m_{h'}^2&=&\frac{-2 \mu^2 \lambda \lambda_\phi -\frac{1}{2} \lambda \lambda_m \mu_\phi^2+\frac{3}{2} \lambda \lambda_m \mu^2_\phi }{\lambda \lambda_\phi -\frac{1}{4}\lambda_m^2} \,, \\
m_{\eta'}^2&=& \frac{-2 \mu_\phi^2 \lambda \lambda_\phi -\frac{1}{2} \lambda_\phi \lambda_m \mu^2+\frac{3}{2} \lambda_\phi \lambda_m \mu^2 }{\lambda \lambda_\phi -\frac{1}{4}\lambda_m^2} \,, \\
m^2_{h'\eta'}&=&\lambda_m v v_\phi \,.
\end{eqnarray}
Diagonalizing the mass matrix with
\begin{equation}
\left(
\begin{array}{ c}
   h'\\
   \eta'
\end{array}
\right) = \left(
\begin{array}{ cc}
  \cos \beta&   \sin \beta \\
   -\sin \beta & \cos \beta  
\end{array}
\right) \,\,
\left(
\begin{array}{ c}
   h\\
   \eta
\end{array}
\right)\,,
\end{equation}
where $\tan 2 \beta= 2 m^2_{h'\eta'}/(m^2_{\eta'}-m^2_{h'})$, one obtains the following Lagrangian in the physical state basis (omitting all SM terms which are unaffected by the $h'$-$\eta'$ mixing)
\begin{eqnarray}
{\cal L}&=& -\frac{1}{4} F_{\mu \nu} \cdot  F^{\mu\nu} 
+\frac{1}{8} (g_\phi v_\phi)^2 A_\mu \cdot A^\mu
+\frac{1}{2}(d_\mu \eta)^2 +
\frac{1}{2}(d_\mu h)^2-\frac{1}{2}m^2_{\eta} \eta^2-\frac{1}{2} m^2_{h} h^2
\nonumber\\
&+&A_\mu \cdot A^\mu [ \kappa^\phi_\eta \eta^2 +\kappa_h^\phi h^2 + \kappa_{h \eta}^\phi \eta h +2 v_\phi \xi^\phi_\eta \eta + 2 v_\phi \xi^\phi_h h] \nonumber\\
&+&(2 W_\mu^+ W^{-\mu} + \frac{1}{\cos^2 \theta_W} Z^\mu Z_\mu) [ \kappa_\eta \eta^2 +\kappa_h h^2 + \kappa_{h \eta} \eta h +2 v \xi_\eta \eta + 2 v \xi_h h] \nonumber\\
&-&\lambda_{\eta}\eta^4 -\lambda_{h} h^4 -\lambda_1 \eta^2 h^2 -\lambda_2 h^3 \eta-\lambda_{3} h \eta^3-\rho_{\eta} \eta^3 - \rho_{h} h^3 - \rho_1 \eta^2 h - \rho_2 h^2 \eta \,.
\label{lagrfin}
\end{eqnarray} 
The various parameters of this Lagrangian are given in appendix A in terms of the input parameters
of Eq.~(\ref{inputlagr}).
Note that, beside the SM parameters, this model has only 4 free parameters what makes it potentially well testable: the four input  parameters $g_\phi$, $\lambda_\phi$, $\mu_\phi$ and $\lambda_m$ or equivalently the masses, symmetry breaking scale and gauge coupling, $m_A$, $m_\eta$, $v_\phi$ and $g_\phi$.

\section{Relic density}

\begin{center}
\begin{figure}[t]
\vspace{0mm}
\hspace{3mm}
\epsfig{file=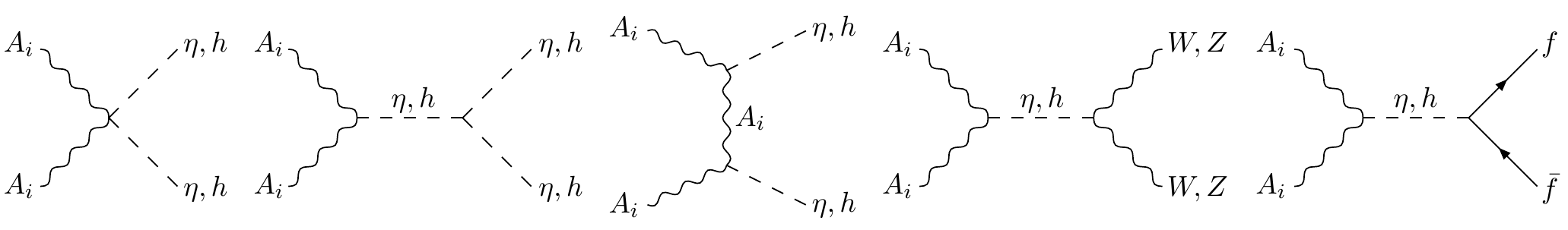,width=14.85cm}
\vspace{-2mm}
\caption{Annihilation diagrams with no DM particle in the final state}
\vspace{-4mm}
\end{figure}
\end{center}
\begin{center}
\begin{figure}
\vspace{5mm}
\hspace{45mm}
\epsfig{file=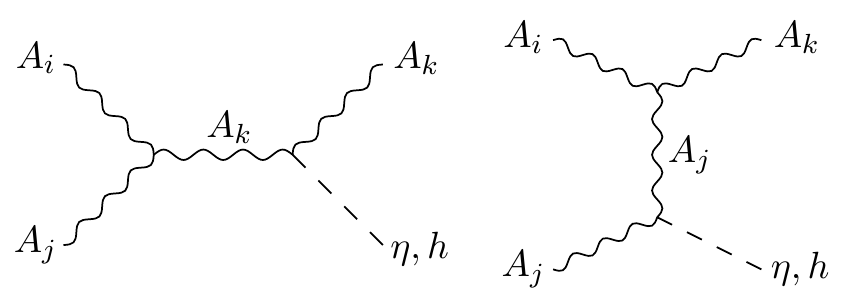,width=5.45cm}
\vspace{-2mm}
\caption{Annihilation diagrams with one DM particle in the final state}
\vspace{-2mm}
\end{figure}
\end{center}
\vspace{-17mm}
There is a number of tree level annihilation processes which determine the vector multiplet relic density through their freezeout. They can be classified in two classes. First the annihilations with no trilinear gauge interactions:
$A_i A_i \rightarrow \eta \eta, \,\eta h,\, h h$ via a direct quartic coupling or via an intermediate $h$ or $\eta$; and
$A_i A_i \rightarrow f \bar{f}, W^+ W^-, Z^0 Z^0$ via a $h$ or $\eta$ in the s-channel, Fig.~1. Second the channels with one trilinear gauge coupling, Fig.~2. These channels have the peculiar property (impossible with ordinary models based on a $Z_2$ symmetry) 
to involve one DM particle in the final state
by reducing the number of DM particles from 2 units to one unit. In the following, at the exploratory level of this article, we calculate the relic density numerically, limiting ourselves 
to the diagrams of Fig.~1 which in most of the parameter space are expected to give a relatively good approximation.\footnote{We leave to a subsequent publication \cite{HT2} the analytic determination of all the cross sections as well as the incorporation of the diagrams of Fig.~2. These diagrams are of same order in $g_{\phi}$ than the first and third diagrams of Fig.~1 and do not cause any particular large effects in the Boltzman equations. This means that in order to get the observed relic density their contribution can be compensated by a moderate decrease of the coupling $g_\phi$ (and therefore a moderate increase of $v_\phi$ for $m_{DM}$ fixed). Similarly the incorporation of these scatterings will not spoil the two operative regime picture explained below. To incorporate these effects necessitates a modification \cite{HT2} of the source code of the program MicroMega2.0 \cite{Belanger:2006is} we have used.}

In the following we will require that the relic density obtained is within the 3$\sigma$ range $0.091 \lesssim \Omega h^2 \lesssim 0.129$ \cite{dunkley}. 

\section{Direct detection}

At tree level a vector DM particle can collide elastically a nucleon either through $h$ exchange or via $\eta$ exchange, which results in a spin independent cross section
\begin{equation}
\sigma_{SI} (N A\rightarrow N A)=\frac{1}{64 \pi} f^2 g_\phi^4 \sin^2 2\beta \,m_N^2\frac{v_\phi^2}{v^2}  \frac{(m_\eta^2-m_h^2)^2}{m_\eta^4 m_h^4} \frac{m_r^2}{m^2_{A}} \,,
\label{directdetection}
\end{equation}
with $m_r=m_N m_{A}/(m_N+m_{A})$ the reduced mass and $m_N$ the nucleon mass. $f$ parametrizes the Higgs nucleon coupling, $f m_N \equiv \langle N | \sum_q m_q \bar{q}q|N\rangle=g_{hNN} v$. We take the value $f=0.3$.
For $m_{A}>>m_N$ numerically one gets:
\begin{equation}
\sigma (N A \rightarrow N A)=1.9 \cdot 10^{-44} \,\hbox{cm}^2 \cdot R\,\sin^2{2\beta}\,\Big(\frac{f}{0.3}\Big)^2 \Big(\frac{g_\phi}{0.5}\Big)^4 \Big( \frac{v_\phi}{500\,\hbox{GeV}}\Big)^2
 \Big(\frac{120\,\hbox{GeV}}{m_h}\Big)^4 \Big(\frac{100\,\hbox{GeV}}{m_A}\Big)^2
\end{equation}
with $R \equiv (m_\eta^2-m_h^2)^2/m_\eta^4$ which is unity for $m_\eta>>m_h$. This has to be compared with the present experimental upper bound on this cross section \cite{CDMS} which, for example around $m_{DM}=100$~GeV, is of order $10^{-44}$~cm$^2$. Therefore for not too small $\beta$ mixing angle the expected signal can be easily of order the present sensitivity or even exceed it by few orders of magnitude. For small mixing angle it can be well below it, independently of the relic density constraint (see below). Direct detection constraints are therefore already relevant to exclude a part of the parameter space.

\section{Electroweak precision measurement constraints}

The main constraint comes from the contribution of the $\eta$ scalar to the $T$ parameter which is the same as the one 
of a scalar singlet mixing with the Higgs boson \cite{Profumo:2007wc,Barger:2007im}:
\begin{eqnarray}
\nonumber
T-T_{SM} & = & \left(\frac{3}{16\pi{\hat s}^2}\right)\Biggl\{ \frac{1}{c^2}\left(\frac{m_H^2}{m_H^2-M_Z^2}\right)\, \ln\, \frac{m_H^2}{M_Z^2} - \left(\frac{m_H^2}{m_H^2-M_W^2}\right)\, \ln\, \frac{m_H^2}{M_W^2}\\
\label{eq:deltaT}
& - & \cos^2\theta\, \left[\frac{1}{c^2}\left(\frac{m_h^2}{m_h^2-M_Z^2}\right)\, \ln\, \frac{m_h^2}{M_Z^2} - \left(\frac{m_h^2}{m_h^2-M_W^2}\right)\, \ln\, \frac{m_h^2}{M_W^2}\right] \nonumber \\
& - & \sin^2\theta\, \left[\frac{1}{c^2}\left(\frac{m_\eta^2}{m_2^2-M_Z^2}\right)\, \ln\, \frac{m_\eta^2}{M_Z^2} - \left(\frac{m_\eta^2}{m_\eta^2-M_W^2}\right)\, \ln\, \frac{m_\eta^2}{M_W^2}\right]\Biggr\}\ \ \ ,
\label{DeltaT}
\end{eqnarray}
where ${\hat s}^2\equiv\sin^2{\hat\theta}_W(M_Z)$ gives the weak mixing angle in the $\overline{MS}$ scheme at the scale $\mu=M_Z$, $c^2=M_W^2/M_Z^2$, and $m_H$ is the reference value of the Higgs boson mass in the SM. The phenomenology of Eq.~(\ref{DeltaT}) has been studied in Ref.~\cite{Profumo:2007wc} (see Figs.~9-10 of this reference where similar expressions for the less important $S$ and $U$ parameters can also be found).
For $m_h=m_\eta$ the constraints on $m_h$ and $m_{\eta}$ are the same as the ones on the Higgs boson mass in the SM: $m_h< 154$~GeV at $95\%$ confidence level \cite{mHbound}. For maximal mixing if one scalar mass is below this value the other one can be larger but should remain under $\sim 250$~GeV. For smaller mixing angle the $\eta'$ dominated scalar can be easily much heavier but the one predominantly composed of $h'$ must remain low. In the following we will limit ourselves to require that $T-T_{SM}$ is within the conservative range $-0.27$-$+0.05$ \cite{Trange} (taking $m_H=114.4$~GeV in Eq.~(\ref{DeltaT})).
For $m_\eta < 114.4$~GeV we will also require that the $\eta\rightarrow f \bar{f}$ branching ratio (or equivalently $\sin^2 \beta$) is below the upper bound from direct search at LEP, Fig.~10 of Ref.~\cite{LEPbbbar}.

\section{Results}

\subsection{Small Higgs portal coupling}

If $\lambda_m$ is small, but large enough to thermalize the $\eta$ and $A_i$'s with the SM thermal bath prior to DM freeze out,  (for instance for example for $m_A\sim 100$-$1000$~GeV, within the range say $\lambda_m\simeq 10^{-8}-10^{-3}$), the $h'$-$\eta'$ mixing angle $\beta$ is small (except for $m_{h'} \simeq m_{\eta'}$ in case the mixing can be large independently of the size of the off-diagonal term in Eq.~(\ref{massmatrix})). 
Except for this case, and except also for $m_h \sim 2 m_A$ or $m_h \sim 2 m_A$ (where the $h$ or $\eta$ exchange diagrams can be resonantly enhanced and therefore be relevant even for small $\beta$ angle), the
only relevant process for the relic density in Fig.~1 is the $A_i A_i \rightarrow \eta \eta$ process.
In this case this process depends only on $g_\phi$, $v_\phi$ and $\lambda_\phi$. If $\lambda_\phi$ is in addition small, only the first and third diagrams of Fig.~1 remains and the annihilation cross section depends only on $g_\phi$ and $m_A$ (or equivalently $v_\phi$). The dependence in the small $\eta$ mass, $m_\eta \simeq \sqrt{ 2 \lambda_\phi} v_\phi$ can be neglected. 
This leads to a cross section proportional to $g_\phi^4/m_A^2$, that is say proportional to $g_\phi^2/v_\phi^2$, and therefore to a linear 
correlation between  $g_\phi$ and $v_\phi$ given in Fig.~3.a (red dots for instance).  Assuming a perturbative $g_\phi$ coupling leads therefore to an upper bound which is of order few tens of TeV depending on the perturbativity condition considered, we get for example $m_A \lesssim 25$~TeV if $g_\phi < 4 \pi$ (in agreement with the unitarity bound which holds for any thermal particle whose relic density results from the freezeout of its annihilation \cite{kamionk}). The corresponding values of $m_\eta$ versus $m_A$ are given in Fig.~3.b. $m_\eta$ scales like $\sqrt{m_A}$ due to the linear correlation between $g_\phi$ and $v_\phi$.
\begin{figure}
\begin{tabular}{c}
\includegraphics[width=7.3cm]{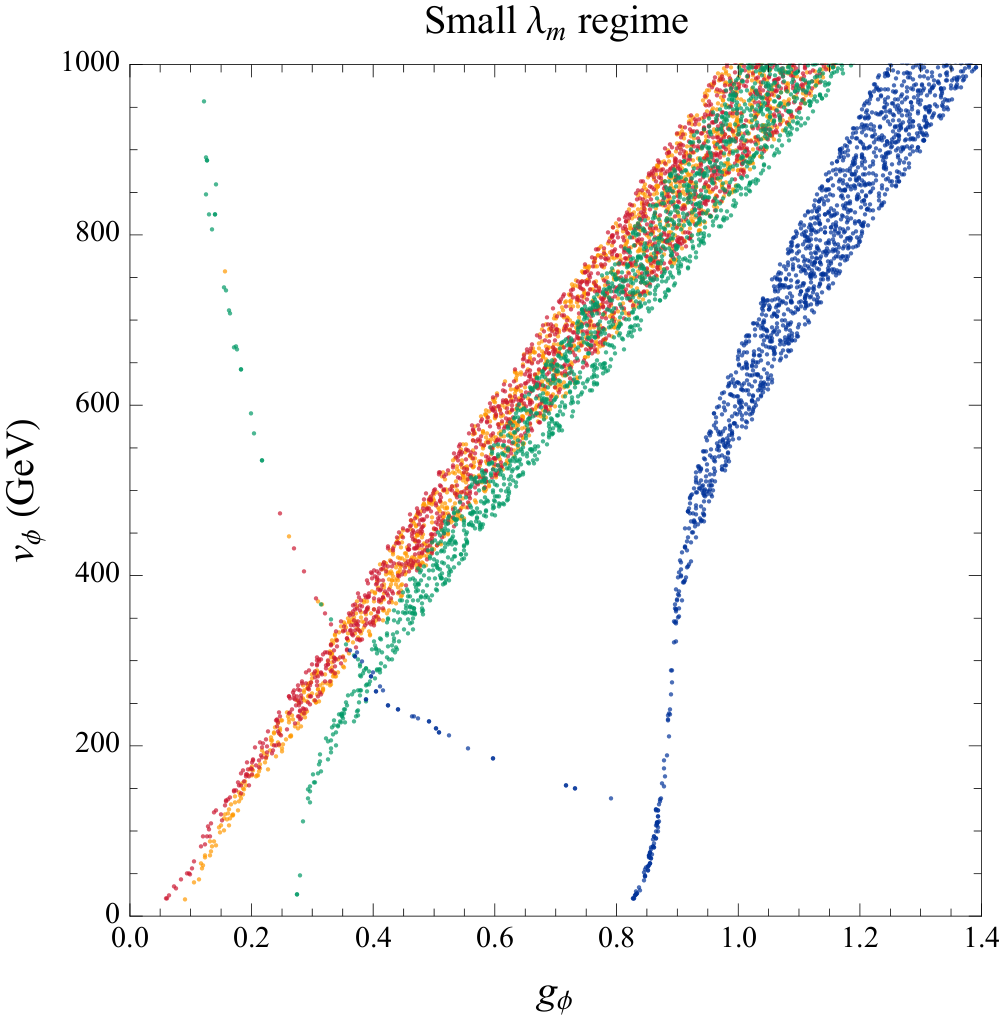}
\includegraphics[width=7.3cm]{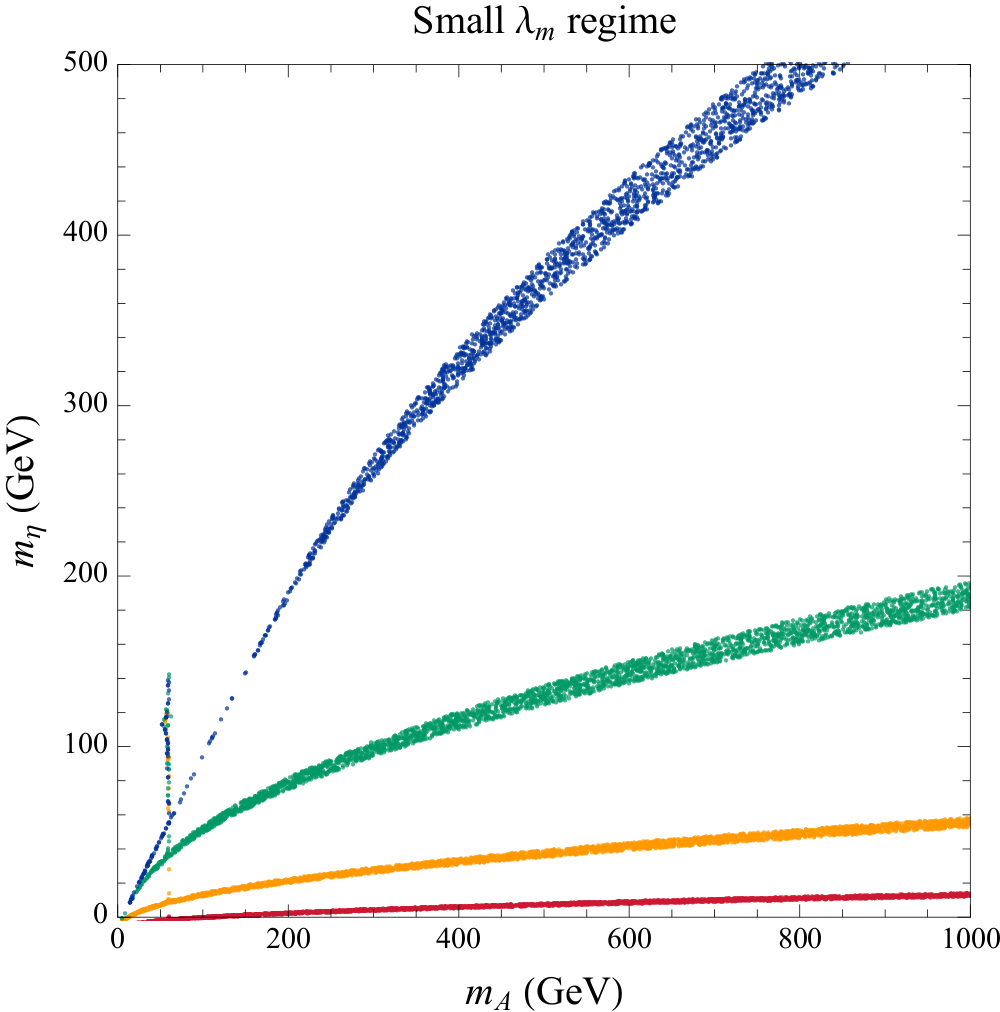}
\end{tabular}
\caption{$g_\phi$ vs $v_\phi$ and $m_A$ vs $m_\eta$ leading to $0.091 \lesssim \Omega h^2 \lesssim 0.129$, for $10^{-7} < \lambda_m < 10^{-3}$, $m_h=120$~GeV and various values of $\lambda_\phi$: $\lambda_\phi= 10^{-4}$ (red), $\lambda_\phi= 10^{-3}$ (orange), $\lambda_\phi= 10^{-2}$ (green) and $\lambda_\phi= 10^{-1}$ (blue) (from left to right and top to bottom respectively). One also recognizes the $m_A =g_\phi v_\phi/2 \sim m_h/2$ resonant case curve.} \label{efficiencies}
\end{figure}
Both $m_{A}$ and $m_\eta$ can be as small as 1~GeV or even much less.\footnote{We don't plot what happens below 1 GeV because the code Micromega we used doesn't allow us to go much below this scale.  But the linear correlation above between $g_\phi$ and $v_\phi$ clearly holds to much lower values as long as $\lambda_\phi$ is small enough. The masses must be nevertheless above the $\simeq$~MeV scale due to BBN constraints \cite{BBN}. 
Whether this could lead to an explanation of the 511~KeV gamma rays from the galactic center observed by Integral \cite{integral} would be worth to be analyzed (through $DM DM \rightarrow \eta \eta$ annihilation followed by $\eta$ decays to $e^+ e^-$).} For low $h$-$\eta$ mixing such low values of the masses are allowed by electroweak data as well as by direct search limits from LEP. The constraints on the Higgs mass, to a good approximation, are the same as in the SM.
As for the direct detection rate, it is proportional to the small Higgs portal interaction $\lambda_m$, and consequently essentially decouples from the relic density which, as shown above, is essentially determined by the pure hidden sector parameters independently of $\lambda_m$.\footnote{This is similar to what happen's in "secluded" DM models \cite{Pospelov}, see also Ref.~\cite{weiner}.} As a result, for $10^{-7}<\lambda_m<10^{-3}$, it is e.g.~few orders of magnitudes below the present upper bounds on the direct detection rates \cite{CDMS}. This decoupling allows to avoid the tension which exists between both constraints in various models. Note however that for $m_\eta << m_h$ (i.e. small $\lambda_\phi$), the cross section of Eq.~(\ref{directdetection}) goes like $m_N^4 g_\phi^2 \lambda_m^2/m_h^4 v_\phi^2 \lambda_\phi^2 \propto \lambda_m^2/\lambda_\phi^2$, so that even for small $\lambda_m$ we can get a large direct detection cross section if $\lambda_\phi$ is even more suppressed. For example for $\lambda_\phi$ as small as $10^{-4}$ and $\lambda_m = 10^{-3}$ we get a cross section which can be as high as $10^{-43}$~cm$^2$ independently of $m_A$. 

For larger value of $\lambda_\phi$ the second diagram of Fig.~1 also becomes important which modifies the correlation between $g_\phi$ and $v_\phi$, Fig.~3. Larger values of $g_\phi$ are necessary for small $v_\phi$. For large value of $\lambda_\phi$ in order to have enough suppression of the annihilation cross section one needs either $m_\eta$ close to $m_A$ or $m_A$ in the multi-TeV range (to benefit from the $1/m^2_A$ asymptotic behavior of the cross sections), or one must have one of the 2 resonances above effective. 
 
\subsection{Large Higgs portal coupling}

For larger values  of $\lambda_m$, say $\lambda_m > 10^{-3}$, the $h$-$\eta$ mixing is larger and annihilations channels other than $AA \rightarrow \eta \eta$ in Fig.~1 become important, or even dominate the DM freezeout process. This leads to a more complex allowed parameter space. In agreement with the electroweak data and LEP constraints above, Fig.~4 displays sets of values of the parameters which lead to a relic density within the WMAP range above as well as to a direct detection cross section below the current upper limits \cite{CDMS}. We find such sets of values for $h$-$\eta$ mixing as large as maximal. For such large mixing both $h$ and $\eta$ masses have to a very good approximation to be above $114.4$~GeV to accommodate the LEP direct search limits and must remain low to accommodate electroweak precision data constraints, see section 5. In this case the LHC experiments should be able to discover both scalars, just in the same way as for the Higgs boson in the SM.
As for the vector bosons they should also be presumably discoverable at LHC through virtual $\eta$, although this will depend on their mass. 
Their mass can lie within a large range, even below the GeV scale in case the relic density can be obtained predominantly from the annihilation to fermions via $h$ and/or $\eta$ exchange. 
\begin{figure}
\begin{tabular}{c}
\includegraphics[width=4.9cm]{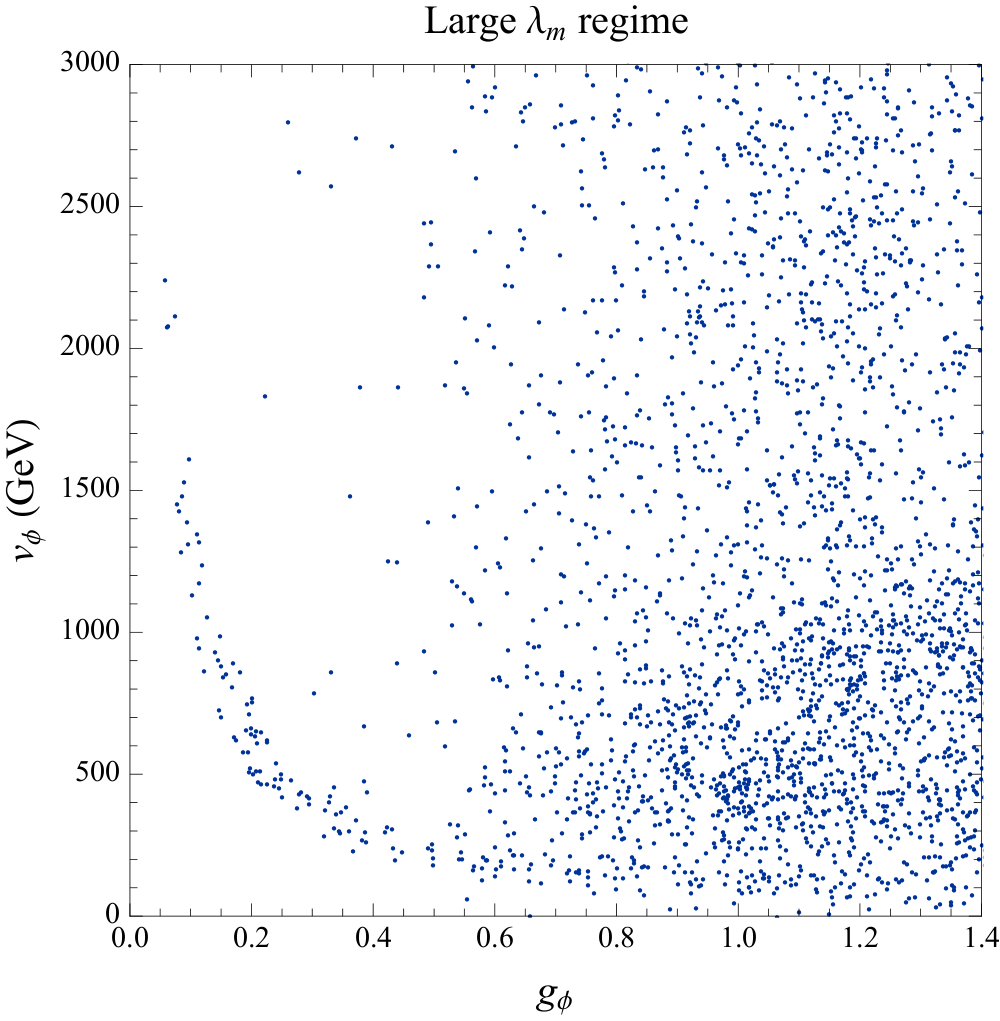}
\includegraphics[width=4.9cm]{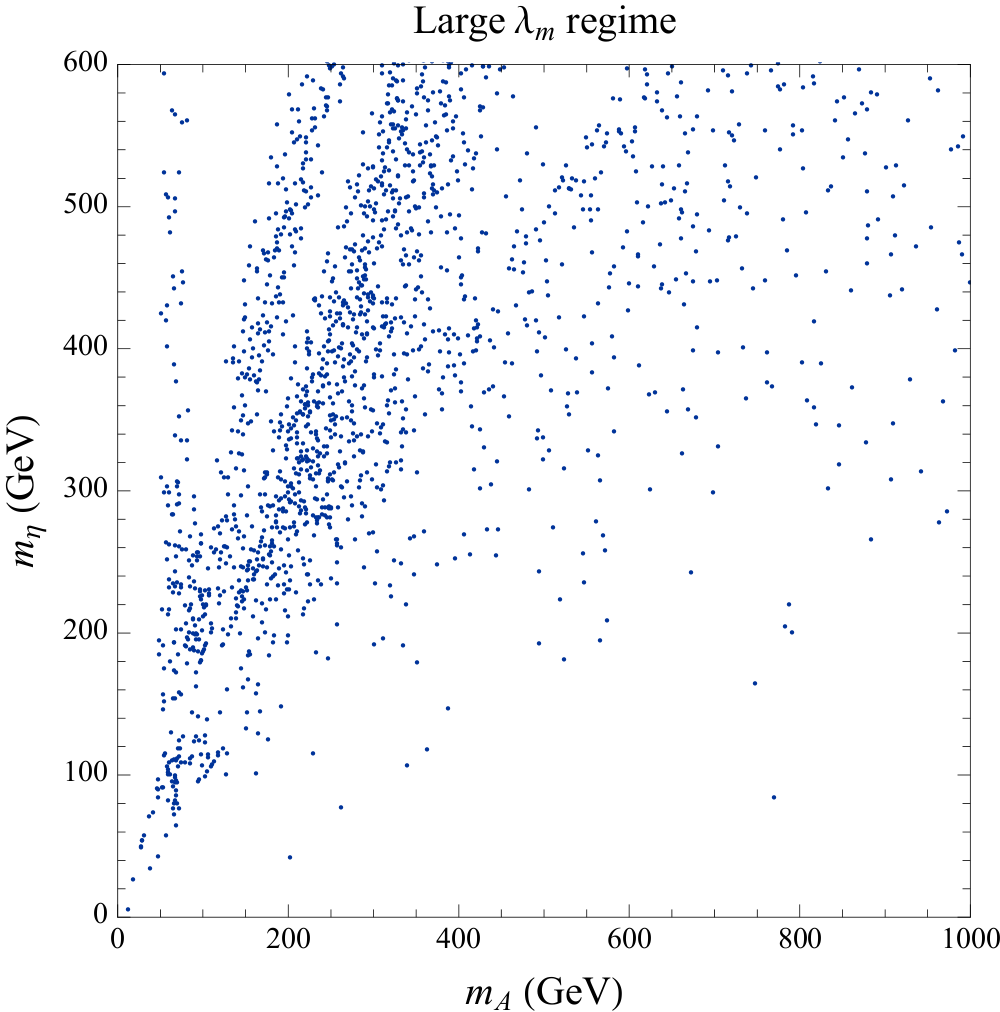}
\includegraphics[width=4.9cm]{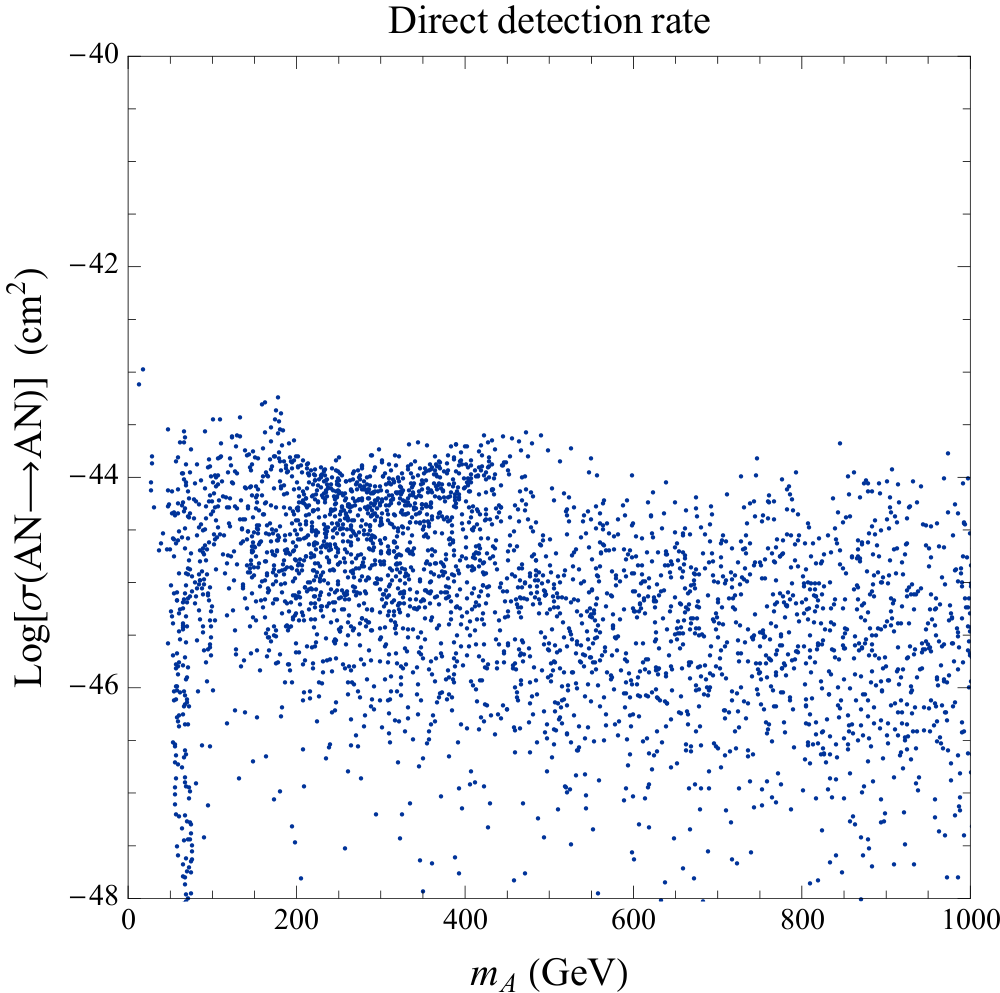}
\end{tabular}
\caption{For $10^{-3} < \lambda_m < 1$ and $114.4\,\hbox{GeV} < m_h < 180~\hbox{GeV}$, values of $g_\phi$ vs $v_\phi$, $m_A$ vs $m_\eta$ and $m_A$  vs $\sigma(AN \rightarrow A N)$ leading to $0.091 \lesssim \Omega h^2 \lesssim 0.129$.
$\lambda_\phi$ has been varied between $10^{-5}$ and $1$. 
Dots with $m_\eta\simeq 2 \, m_A$ proceed through resonance of the $\eta$ exchange diagrams (to $W^+W^-$, $ZZ$, $f \bar{f}$ or $hh$). Similarly dots with $m_h\simeq2 \, m_A$ similarly are dominated by the Higgs exchange diagrams (to $W^+W^-$, $ZZ$, $f \bar{f}$ or $\eta \eta$). Dots with   $m_\eta <114.4$~GeV are for suppressed values of $\sin \beta$ to agree with the LEP constraints on the $h \rightarrow f \bar{f}$ branching ratio.} 
\end{figure}

As for the direct detection rate, we obtain a large range of cross section values, from several orders of magnitude below the current experimental limits to several orders of magnitude above them. Fig.~4 gives the sets of values of parameters which are consistent with these current limits. 
It would be worth to see if this model could also lead to interesting DM indirect detection signals (which, as for direct detection, are proportional to $\sin^2 \beta$ and therefore decouple from the relic density constraints for small mixing angle).
In particular for what concerns positron indirect detection, as the $\eta$ or $h$ scalar mediated interaction between 2 gauge bosons is attractive, it can lead to an enhancement of the positron emission through Sommerfeld effects \cite{sommerfeld,cirelli3}. It is to be analyzed in detail if, without applying large ad hoc boost factors, this enhancement can be large enough to explain the excess of positrons observed by the Pamela experiment \cite{pamela}. From the results of Ref.~\cite{cirelli3}, Fig.~10 in particular, this appears to be possible, either for $m_A$ in the multi TeV range through dominant annihilation to $hh$, $h \eta$, $\eta \eta$, or for smaller values of $m_A$ (but larger than $\sim$~100 GeV) with $m_\eta$ around $\sim 1~$GeV.
In the latter case, $\eta$ particles (from $A_i A_i \rightarrow \eta \eta,h\eta$ annihilations) decay predominantly to leptons (leading to a hard enough positron spectrum) and serve as mediators to produce highly non-relativistic leptons in large quantity, as in the mediator mechanism very recently proposed in Ref.~\cite{arkani}. Like in many models with such a light mediator this mechanism suffers nevertheless from stability problem of the light mediator mass under radiative corrections (i.e.~in our case the stability of the $\eta$ mass under gauge induced self-energy corrections).

\section{Effects of higher dimensional operators}

The model above doesn't necessarily requires a UV completion.\footnote{In particular as it involves only one gauge interaction in the hidden sector and no fermions it doesn't call for any particular grand-unification UV completion in the hidden sector. As in the SM, there is though a hierarchy problem related to the stabilization of the mass of the Higgs boson $\eta$ (under radiative corrections from any new physics or from the gravitation). We do not address this problem.} A question one must ask nevertheless is, if there exist heavier particles in the hidden sector at a higher energy scale $\Lambda$, how these particles could spoil the stability of the vector bosons.  If these particles induce dimension five operator destabilizing the vector bosons, one expects a lifetime, $\tau_A \sim c \,\Lambda^2/m_A^3$ (with $c$ a coefficient of order $4\pi$ for a two body decay), many orders of magnitude smaller than the age of the universe, even for $\Lambda$ as high as the GUT scale (unless the involved couplings are highly suppressed).
However in the hidden vector model above there is no gauge invariant dimension 5 operators which could be induced by a higher energy physics. Only dimension 6 operators can be induced, which induce much longer lifetimes, see e.g.~\cite{Cirelli:2005uq}. 

For example one gauge invariant operator which could be induced is 
$D_\mu \phi^\dagger \phi D^\mu H^\dagger H/\Lambda^2$.
Since $\phi$ is assumed to be at a low scale this operator is relevant.  
It induces for example the decay $A_i\rightarrow \phi \phi^*$ with a lifetime of order $\tau_A \sim 4 \pi \Lambda^4/m_A^5$ which for $m_A\simeq 1$~TeV ($\simeq 1$~GeV) is longer than the age of the universe if $\Lambda$ is above $\sim 10^{13}$~GeV ($10^{9}$~GeV), or less if the involved couplings are smaller  
than unity. This means that the heavy particles at the origin of these operators must be heavier than these scales (just as for the proton in the SM in presence of any $B$ violating new physics source, but at a much less constrained level because the lower limit on the proton lifetime is far larger than the age of the universe). 
Note that there are other dimension six operators which could cause the decay of the vector bosons, with similar lower bound on the underlying scale, $\phi^\dagger F^i_{\mu\nu} \frac{\tau_i}{2} \phi F^{Y\mu\nu}/\Lambda^2$,
$D_\mu \phi^\dagger \phi H^\dagger D^\mu H/\Lambda^2$, $D_\mu \phi^\dagger D_\nu \phi F^{Y\mu\nu}/\Lambda^2$.
There are also dimension 6 operators which do not violate the custodial symmetry, therefore not causing any decay,
$D_\mu \phi^\dagger F^i_{\mu \nu} \frac{\tau_i}{2} D^\nu \phi/\Lambda^2$, $D^\mu \phi^\dagger D_\mu \phi H^\dagger H/\Lambda^2$ and 
$F^i_{\mu \nu} F^{i\mu \rho} F^{Y\nu}_{\rho}/\Lambda^2$.\footnote{Note also that if one 
adds in the hidden sector light fermions
to which the hidden vector can decay, still one would get 
a DM candidate: the lightest fermionic multiplet of the non-abelian gauge-symmetry (and SM singlet) would provide a stable DM candidate because such hidden sector fermions can couple only in pairs to any other particle. This shows how generically gauged sectors coupling to the SM through the Higgs portal can provide DM candidates. It is beyond the scope of this work to study such possibilities which contain more fields than the hidden vector model, and are very different in many respects.} 

It has been shown in Ref.~\cite{Sannino} that DM decaying dominantly to $e^+ e^-$ pairs with lifetime of order $\sim 10^7$ times the age of the universe, which we typically obtain from a dimension 6 operator with $\Lambda$ of order the GUT scale,
leads to a positron flux with the right order of magnitude to explain the excess of positrons
observed by the Pamela experiment. This leads also to a cosmic $\gamma$ flux that can be probed by current experiments.\footnote{Possible consequences of the decays induced by the operators above, in particular emission of intense $\gamma$ lines, and excess of positrons, are analysed in Ref.~\cite{Arina:2009uq}.}

\section{A few more comments}

If one adds to the SM as few new fields as possible there are not that many possibilities to obtain a stable DM candidate without assuming by hand a discrete or global symmetry.

One possibility, which holds with only one extra field \cite{Cirelli:2005uq}, assumes the existence of a high fermion or scalar $SU(2)_L$ multiplet: a fermion quintuplet or higher or a scalar sextuplet or higher.
Such multiplets are stable because no gauge invariant operators destabilizing these multiplets can be written with dimension less than 6.

Another possibility, which holds with 2 extra fields, is the hidden vector model above. It involves lower multiplets. 

A third possibility, which holds with 3 extra fields, has been proposed in Ref.~\cite{Pospelov}. It assumes a $U(1)'$ gauge boson, a scalar charged under the $U(1)'$ and a fermion also charged under it. The scalar breaks the $U(1)'$ to make the gauge boson massive and the fermion is the stable DM particle. The SM and hidden sector can communicate through both kinetic mixing and Higgs portal interactions.\footnote{Note that, with many more fields, mirror models which consider a complete copy of the SM in the hidden sector, can also lead to stable candidates (in particular the mirror proton) without assuming a discrete or global symmetry by hand, see e.g.~Ref.~\cite{foot}.}

Finally note also that if in the hidden vector model above one considers an abelian gauge group $G'=U(1)'$ (instead of the $SU(2)_{HS}$ above), with $\phi$ a complex scalar charged under it (instead of the doublet above), one ends up with a Lagrangian similar to the one of Eq.~(\ref{lagrfin}),
with nevertheless 3 important differences. First, by replacing the non-abelian field in Eq.~(\ref{lagrfin}) by the abelian one, the trilinear gauge couplings disappear and therefore all annihilation processes of Fig.~2 do not exist, but these processes are not mandatory to obtain the experimental relic density. 
Second with an abelian gauge symmetry there is no more custodial symmetry to make the gauge boson stable but still the Lagrangian of Eq.~(\ref{lagrfin}) with an abelian field instead of the non-abelian one displays a $Z_2$ symmetry (under which $A_\mu$ is odd with all other fields even) related to the charge conjugation symmetry of the Lagrangian (which results from the gauge symmetry and particle content). Third, but not least, in order that the $U(1)'$ gauge boson is stable one has to make the assumption that there is no $F'_{\mu \nu} F^{\mu \nu}_Y$ kinetic mixing interaction with the hypercharge gauge boson.
Unlike in the non-abelian case this term is not forbidden by any symmetry of the model. This cannot be justified without assuming extra symmetries. 
But it can be noted that, would this kinetic mixing term be absent, would all numerical results obtained above hold also for this case apart from factors of order unity (for instance a factor $1/3$ in the relic density because there is only one DM component instead of 3).\footnote{The phenomenology of a similar model has been studied in Ref.~\cite{jwu} independently of DM considerations.}

\section{Summary}

We have shown that a hidden sector vector multiplet associated to a non-abelian gauge group $G'$, coupling to the SM only through the Higgs portal interaction of a scalar charged under this gauge group, constitutes a perfectly viable DM candidate. This vector multiplet is stable without needing to assume any discrete or global symmetry, due to the custodial symmetry of the Lagrangian which results from the gauge symmetry.
The stability of the gauge bosons could be spoiled by physics at higher energies but only through dimension six operators which is fine as long as the new physics is above a high scale, $\sim10^{14}$~GeV ($\sim10^{10}$~GeV) for a DM mass equal to 1~TeV (1~GeV). For small Higgs portal coupling (but large enough to thermalize the hidden sector with the SM particles prior to freeze-out) the relic density is determined only by hidden sector parameters, while the direct detection rate necessarily involves the Higgs portal quartic coupling. This allows to decouple the direct detection constraints from the relic density ones. Large Higgs portal interaction, i.e. large mixing between the Higgs boson and the extra scalar, is also allowed in a large fraction of the parameter space, in case the model is testable at accelerators. The gauge boson mass can lie within a wide range of values from $\sim$~MeV to few tens of TeV. This model can also lead to a Sommerfeld enhancement of the positron emission relevant for the recent Pamela experiment result.

\section*{Acknowledgments}
We specially thank M.~Papucci who participated in an early stage of this work, and M.~Tytgat for discussions. We also acknowledge interesting discussions with J.-M. Fr\`ere, F.S.~Ling, L.~Lopez-Honorez, G.~Servant and G.~Vertongen. This work is supported by the FNRS-FRS and the Belgian Science Policy (IAP VI-11), IISN.

\appendix
\section{Appendix}

In terms of the various input parameters of Eq.~(\ref{inputlagr}), the parameters of Eq.~(\ref{lagrfin}) read as follows:
\begin{eqnarray}
m^2_h&=&m^2_{h'} \cos^2 \beta + m^2_{\eta'}\sin^2 \beta  - m^2_{h'\eta'} \sin 2 \beta \nonumber \\
m^2_\eta&=&m^2_{\eta'} \cos^2 \beta + m^2_{h'}  \sin^2 \beta +m^2_{h'\eta'} \sin 2 \beta \nonumber \\
\kappa_\eta^\phi&=&\frac{1}{8} g_\phi^2 \cos^2 \beta\nonumber\\
\kappa_h^\phi&=&\frac{1}{8} g_\phi^2 \sin^2 \beta\nonumber\\
\kappa_{h\eta}^\phi&=&-\frac{1}{8} g_\phi^2 \sin 2 \beta\nonumber\\
\kappa_\eta&=&\frac{1}{8} g^2 \sin^2 \beta\nonumber\\
\kappa_h&=&\frac{1}{8} g^2 \cos^2 \beta\nonumber\\
\kappa_{h \eta}&=&\frac{1}{8} g^2 \sin 2 \beta\nonumber\\
\xi^\phi_\eta&=&\frac{1}{8} g^2_\phi \cos \beta\nonumber \\
\xi^\phi_h&=&-\frac{1}{8} g^2_\phi \sin \beta\nonumber \\
\xi_\eta&=&\frac{1}{8} g^2 \sin \beta\nonumber \\
\xi_h&=&\frac{1}{8} g^2 \cos \beta\nonumber \\
\lambda_\eta&=&\frac{1}{4} (\lambda_\phi \cos^4 \beta + \lambda \sin^4 \beta +\lambda_m \cos^2 \beta \sin^2 \beta)\nonumber \\
\lambda_h&=&\frac{1}{4} (\lambda_\phi \sin^4 \beta + \lambda \cos^4 \beta +\lambda_m \cos^2 \beta \sin^2 \beta)\nonumber \\
\lambda_1&=&\frac{1}{4} (6 \lambda_\phi \cos^2 \beta \sin^2 \beta + 6 \lambda \sin^2 \beta \cos^2  \beta +\lambda_m (\cos^4 \beta+ sin^4 \beta -4 \cos^2 \beta \sin^2 \beta))\nonumber \\
\lambda_2&=&\frac{1}{4} (-4\lambda_\phi \sin^3 \beta \cos \beta + 4 \lambda \cos^3 \beta \sin \beta -\lambda_m \sin 2 \beta \cos 2 \beta)\nonumber \\
\lambda_3&=&\frac{1}{4} (-4\lambda_\phi \cos^3 \beta \sin \beta + 4 \lambda \sin^3 \beta \cos \beta +\lambda_m \sin 2 \beta \cos 2 \beta)\nonumber \\
\rho_\eta&=&\frac{1}{4}(4 \lambda_\phi v_\phi \cos^3 \beta  + 4 \lambda v \sin^3 \beta  + 2 \lambda_m (v \cos^2 \beta \sin \beta  + v_\phi \sin^2 \beta \cos \beta ))\nonumber\\
\rho_h&=&\frac{1}{4}(-4 \lambda_\phi v_\phi \sin^3 \beta  + 4 v \lambda \cos^3 \beta  + 2 \lambda_m (v \sin^2 \beta \cos \beta  - v_\phi  \cos^2 \beta \sin \beta ))\nonumber\\
\rho_1&=&\frac{1}{4}(-12 v_\phi \lambda_\phi \cos^2 \beta \sin \beta  + 12 \lambda v \sin^2 \beta \cos \beta  +  \lambda_m (2 v \cos^3 \beta  - 2 v_\phi \sin^3 \beta  -4 v \cos \beta \sin^2 \beta  + 4 v_\phi \cos^2 \beta \sin \beta ))\nonumber\\
\rho_2&=&\frac{1}{4}(12  \lambda_\phi v_\phi  \sin^2 \beta \cos \beta  + 12 \lambda v \cos^2 \beta \sin \beta  +  \lambda_m (2 v \sin^3 \beta  + 2 v_\phi \cos^3 \beta   - 4 v \cos^2 \beta \sin \beta  - 4 v_\phi \sin^2 \beta \cos \beta ))\nonumber
\end{eqnarray}


\end{document}